\documentclass[12pt]{iopart}

\usepackage{graphicx} 
\usepackage{amssymb}
\usepackage{url}

\begin{document}

\title{ The Morphospace of Consciousness } 

\author{Xerxes D. Arsiwalla$^{1,2,3}$, Ricard Sol{\'e}$^{4,5,6,7}$,  Cl\'{e}ment Moulin-Frier$^3$,  Ivan Herreros$^3$,  Mart\'i S\'anchez-Fibla$^3$,  Paul Verschure$^{1,2,7}$}

\address{$^1$Synthetic Perceptive Emotive and Cognitive Systems Lab, Institute for BioEngineering of Catalonia (IBEC),  Barcelona, Spain}
\address{$^2$Barcelona Institute for Science and Technology (BIST)}
\address{$^3$Dept. of Information Technologies, Universitat Pompeu Fabra (UPF),  Barcelona, Spain} 
\address{$^4$Complex Systems Lab, Universitat Pompeu Fabra, Barcelona, Spain} 
\address{$^5$Institut de Biologia Evolutiva (CSIC-UPF), Barcelona, Spain}
\address{$^6$Santa Fe Institute, 399 Hyde Park Road, Santa Fe, NM, 87501, USA} 
\address{$^7$Instituci\'{o} Catalana de Recerca i Estudis Avan\c{c}ats (ICREA), 
Barcelona, Spain. }  
\ead{x.d.arsiwalla@gmail.com}
\vspace{10pt}

\begin{abstract}  
In this paper, we construct a complexity-based morphospace wherein one can study systems-level properties of conscious and intelligent systems based on information-theoretic measures. The axes of this space labels three distinct complexity types, necessary to classify conscious machines, namely, autonomous, cognitive and social complexity. In particular, we use this morphospace to compare biologically conscious agents ranging from bacteria, bees, C. elegans, primates and humans with artificially intelligence systems such as deep networks, multi-agent systems, social robots, AI applications such as Siri and computational systems as Watson. Given recent proposals to synthesize consciousness, a generic complexity-based conceptualization provides a useful framework for identifying defining features of distinct classes of conscious and synthetic systems. Based on current clinical scales of consciousness that measure cognitive awareness and wakefulness, this article takes a perspective on how contemporary artificially intelligent machines and synthetically engineered life forms would measure on these scales. It turns out that awareness and wakefulness can be associated to computational and autonomous complexity respectively. Subsequently, building on insights from cognitive robotics, we examine the function that consciousness serves, and argue the role of consciousness as an evolutionary game-theoretic strategy. This makes the case for a third type of complexity necessary for describing consciousness, namely, social complexity. Having identified these complexity types, allows for a representation of both, biological and synthetic systems in a common morphospace. A consequence of this classification is a taxonomy of possible conscious machines. In particular, we identify four types of consciousness, based on embodiment: (i) biological consciousness, (ii) synthetic consciousness, (iii) group consciousness (resulting from group interactions), and (iv) simulated consciousness (embodied by virtual agents within a simulated reality). This taxonomy helps in the investigation of comparative signatures of consciousness across domains, in order to highlight design principles necessary to engineer conscious machines. This is particularly relevant in the light of recent developments at the crossroads of cognitive neuroscience, biomedical engineering, artificial intelligence and biomimetics.

\end{abstract}

%
\vspace{2pc}
  
%
%
%

\section{Introduction  }
Can one construct a taxonomy of consciousness based on evidence from clinical  neuroscience, synthetic biology, artificial intelligence (AI) and cognitive robotics?  In this paper we explore current biologically motivated metrics of consciousness. In view of these metrics, we show how contemporary AI and synthetic systems  measure on homologous scales. In what follows, we refer to a phenomenological description of consciousness. In other words, that which can be described in epistemically objective terms, even though aspects of the problem of consciousness may require an ontologically subjective description. Drawing from what is known about the phenomenology of consciousness in  biological systems, we build a homologous argument for artificial, collective  and simulated systems.  For example, in clinical diagnosis of disorders of consciousness, two widely used scales are patient awareness and wakefulness (also referred to as arousal), both of which can be assessed using neurophysiological recordings  \cite{laureys2004}, \cite{laureys2005}.   We will  use these scales to  construct a {\em morphospace  of consciousness}.  

The origin of the concept of a morphospace comes from comparative anatomy and paleobiology, where either quantitative measures or principal components from a clustering methods allow locating given items in a metric-like space, but it can also involve a  relative position approach, as the one we will follow here. A related concept of the so-called {\em theoretical morphospace}, has also been defined in formal terms, as an $N$-dimensional geometric hyperspace produced by systematically varying the parameter values associated to a given (usually geometric) set of traits \cite{mcghee1999theoretical}.  More recently,  morphospaces have been used in the study of complex systems, linguistics and biology  \cite{avena2015network},   \cite{olle2016morphospace},  \cite{seoane2018morphospace}.  A morphospace commits one to embodiment or form. In the context of consciousness, embodiment can be both, physical and virtual.  Hence, a   morphospace serves as a useful tool to gain insights on design principles and evolutionary constraints,  when looking across a large class of  systems (or species) that display complex variations in traits. For the problem of consciousness, we construct this morphospace based on three distinct complexity types. These considerations suggest  an embodiment-based taxonomy of  consciousness  \cite{arsiwalla2016three}.  


For practical reasons, many experimental paradigms testing consciousness are designed for humans or higher-order primates (see  \cite{baars2005global},  \cite{koch2016neural}, \cite{tononi2016integrated}  for an overview of the field).   In this article, we argue that metrics commonly associated to biological consciousness can  also be meaningfully used for conceptualizing behaviors of synthetic and artificially intelligent systems.   This is insightful not only for understanding parallels between biological and potential synthetic consciousness, but more importantly for unearthing design principles necessary for building biomimetic technology that could potentially acquire consciousness.  As evidenced by several historical precedents, bio-inspired design thinking has been at the core of some of the greatest scientific breakthroughs. For instance, early attempts at aviation in the 19th century were   inspired by studying flight mechanics in birds and insects (the term aviation itself is derived from the Latin "avis" for "bird"). In fact,  biological flight mechanisms are so sophisticated  that their biomimetic  implementations are still being actively studied within the field of soft robotics \cite{mischiati2015}. However, it so happened that rather than coming around to mimicking nature exactly,  humanity learnt the basic laws of aerodynamics based on observations  from nature and looked for other embodiments of those  principles. This in fact, led to the invention of the modern aircraft by the Wright brothers in 1903, leading to a completely new way to build machines that fly than those that exactly mimic nature.  

Another paradigm-changing example of bio-inspired thinking leading to modern day technological innovation can be seen in artificial neural networks, which dates back to the 1930s with the first model of neural networks by Nicolas Rashevsky      \cite{rashevsky1933}, followed by the seminal work of Walter Pitts and Warren McCulloch in 1943  \cite{mcculloch1943logical}  and Frank Rosenblatt's perceptron in 1958  \cite{rosenblatt1958perceptron}.  The field began  as a modest attempt to understand cognition and brain function. Eventually, with the use of analytical tools from statistical physics, those simple formal models paved the way to understanding associative memory and other emergent cognitive  phenomena \cite{hopfield1982neural}. Even though artificial neural networks did not quite solve the problem of how the brain works, they led to the discovery of brain-inspired computing technologies such as deep learning systems and powerful technologies for computational intelligence such as IBM's Watson. These machines process massive volumes of data and are built for intensive computational tasks that the brain is not even designed for.  In that spirit, the next frontier is understanding the governing principles of biological    consciousness and  its various embodiments, which could potentially lead to the growth of next-generation sentient  technologies.  Recent work in this direction can be found in  \cite{sole2017rise}.  

  
Metrics of consciousness are also the right tools to quantitatively study how human intelligence differs from current machine intelligence.  Once again, it is instructive to take a historical perspective on human intelligence as laid out by one of the founders of AI, Allen Newell in 1994 in his seminal work,  "Unified Theories of Cognition" \cite{newell1994unified}.   Newell proposed the following thirteen  criteria necessary for building human-level cognitive architectures:  \\  
$\bullet$ Behave flexibly as a function of the environment  \\
$\bullet$ Exhibit adaptive (rational, goal-oriented) behavior \\  
$\bullet$ Operate in real-time  \\ 
$\bullet$ Operate in a rich, complex, detailed environment (that is, perceive an immense amount of changing detail, use vast amounts of knowledge, and control a motor system of many degrees of freedom)  \\  
$\bullet$ Use symbols and abstractions  \\  
$\bullet$ Use language, both natural and artificial  \\  
$\bullet$ Learn from the environment and from experience  \\  
$\bullet$ Acquire capabilities through development  \\  
$\bullet$ Operate autonomously, but within a social community  \\  
$\bullet$ Be self-aware and have a sense of self   \\ 
$\bullet$  Be realizable as a neural system  \\  
$\bullet$ Be constructible by an embryological growth process  \\  
$\bullet$ Arise through evolution  \\
Current AI architectures still do not meet all these criteria. On the other hand, though Newell did not discuss consciousness back then, the above criteria are very relevant in the light of current research on neural mechanisms of  consciousness \cite{koch2016neural}.   While Newell's   criteria  list  signatures that are the consequence of  human  intelligence,  for consciousness it is more useful to have a list of  functional criteria that results in the process of consciousness.  In this article, we shall discuss  prospective  functional criteria for consciousness.

\section{Biological Consciousness: Insights from Clinical Neuroscience  }

We begin this discussion reviewing clinical scales used for assessing   consciousness in patients with disorders of consciousness. In subsequent sections, we generalize complexity measures pertinent to these biological scales and discuss how current synthetic systems measure up on these. 

\subsection{Clinical Consciousness and its Disorders }

In patients with disorders of consciousness ranging from coma, locked-in syndrome to those in vegetative states, levels of consciousness are assessed through a battery of behavioral tests as well as physiological recordings.  Cognitive awareness in patients  is assessed by testing several cognitive functionalities using behavioral and neurophysiological (fMRI or EEG) protocols \cite{laureys2004}.  Assessments of wakefulness/arousal in patients are based on metabolic markers (in cases where reporting is not possible) such as glucose uptake in the brain, captured using PET scans.  More generally,  in  \cite{laureys2004} and \cite{laureys2005}, awareness and wakefulness have been proposed as a two dimensional operational definition of clinical consciousness,    shown in figure \ref{fig1} below.  While awareness concerns higher and lower-order cognitive functions enabling complex behavior;  wakefulness results from biochemical homeostatic mechanisms regulating survival drives and is clinically measured in terms of glucose metabolism in the brain. In fact, in all known organic life forms, biochemical arousal is a necessary precursor supporting the hardware necessary for cognition. In turn, evolution has shaped cognition in such a way so as to support the organism's basic survival (using wakefulness/arousal) as well as higher-order drives (using awareness) associated to cooperation and competition in a multi-agent environment \cite{verschure2014and}. Awareness and wakefulness thus taken together, form the clinical markers of consciousness.      

\begin{figure}[h]
\centering
\includegraphics[width=0.7\linewidth]{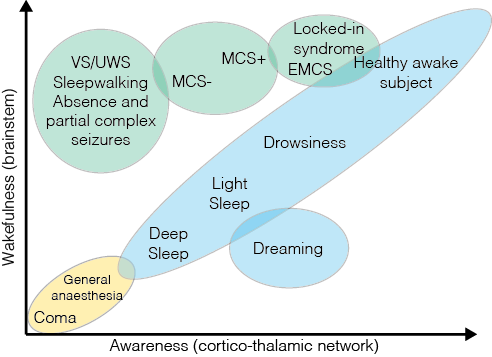}
\caption{{\bf  Clinical scales of consciousness.}  A clustering of  disorders of consciousness in humans represented on scales of awareness and wakefulness.   Adapted from  \cite{laureys2005}. In neurophysiological recordings, signatures of awareness have been found in cortico-thalamic activity, whereas wakefulness corresponds to activity in the brainstem and associated systems  \cite{laureys2004},   \cite{laureys2005}.  Abbreviated legends: VS/UWS (vegetative state/unresponsive wakefulness state)  \cite{laureys2010unresponsive};  MCS(+/-) (minimally conscious state plus/minus), EMCS (emergence from minimally conscious state)    \cite{bruno2011unresponsive}.   }
\label{fig1}
\end{figure}

This clinical definition of consciousness enables a practical classification of closely associated states/disorders of consciousness into clusters on a bivariate scale with awareness and wakefulness on orthogonal axes.  Under healthy conditions, these two levels are almost linearly correlated, as in conscious wakefulness (high arousal and high awareness) or in deep sleep (low arousal and low awareness). In pathological states, wakefulness without awareness can be observed in the vegetative state \cite{laureys2004}, while transiently reduced awareness is  observed following seizures \cite{blumenfeld2012impaired}. Patients in the minimally conscious state show intermittent and limited non-reflexive and purposeful behavior \cite{giacino2002minimally}, \cite{giacino2004vegetative}, whereas patients with hemispatial neglect display reduced awareness of stimuli contralateral to the side where brain damage has occurred \cite{parton2004hemispatial}.  

 
The question is how can one generalize  wakefulness/arousal and awareness for non-biological systems in order to obtain homologous scales of consciousness that can be mapped to artificial systems?  As noted above, wakefulness/arousal results from autonomous homeostatic mechanisms necessary for the self-preservation of an organism's germ line in a given environment. In other words, arousal results from self-sustaining life processes necessary for basic survival, whereas awareness refers to functionalities pertaining to estimating or predicting states of the world and optimizing the agent's own actions with respect to those states.  If biological consciousness as we know it, is a synergy between metabolic and cognitive processes, the question one can ask is how should this insight be extended to conceive a functional notion of consciousness in synthetic systems? One way of doing so might be generalizing wakefulness/arousal to scales of autonomous functioning and awareness to scales of computational or informational  processes.  

\subsection{Measures of Consciousness }

Specific measures of autonomy  and computation/information processing have  been discussed in psychometric \cite{weinstein2012index}  respectively neurophysiological studies \cite{wibral2014directed}. However, applying these measures to artificial systems and comparing those values to biological systems is not always so straightforward (due to completely different processing substrates). Nonetheless, these measures offer a first step in this direction. For example,  \cite{weinstein2012index} introduce an "Index of Autonomous Functioning", tested on healthy human subjects (via psychometric questionnaires). This index aims to assess psychological ownership, interest-taking and susceptibility to external controls. This is similar to the concept of volition (or agency), introduced  in the cognitive neurosciences  \cite{haggard2008human}, which seeks to determine the neural correlates of  self-regulation, referring to actions regulated by internal drives rather than exclusively driven by external contingencies.  

Attempts to quantify awareness have appeared in \cite{deci2000and}, discussed in the context of a unified psychological theory of self-functioning. However, in consciousness research, a measure of awareness that has gained a lot of traction is integrated information  \cite{tononi1994} (often denoted as $\Phi$). This is an information-theoretic complexity measure. It was first introduced in neuroscience as a  measure applicable to neural networks.  Based on mutual information, $\Phi$ has been touted as a  correlate of consciousness  \cite{tononi1994}. Integrated information   is loosely defined as the quantity of information generated by a network as a whole,  due to its causal dynamical interactions,  over and above the information generated independently by the disjoint sum of its parts.  As a complexity measure, $\Phi$  seeks to operationalize the intuition that complexity arises from simultaneous integration and differentiation of the network's structure and dynamics, thus enabling the emergence of the system's collective states.  The interplay between  integration and differentiation  generates information that is highly diversified yet integrated, creating patterns of high complexity. Following initial proposals \cite{tononi2004}, \cite{tononi2003}, \cite{tononi1994}, several approaches have been developed to compute integrated information \cite{arsiwalla2013iit}, \cite{arsiwalla2016computing}, \cite{arsiwalla2016high}, \cite{xda2016global},  \cite{arsiwalla2017brain},  \cite{arsiwalla2018measuring},   \cite{ay2015information}, \cite{bt}, \cite{bs}, \cite{griffith2014principled},  \cite{oizumi2014},  \cite{petersen2015},    \cite{tegmark2016improved},  \cite{wennekers2005stochastic}. 

Notably the work of   \cite{xda2016global} is of particular significance in the context of this discussion  as it develops  large-scale network computations of integrated information, applied  to the human brain's connectome data. The human connectome  data consists of structural connectivity of white matter fiber tracts in the   cerebral cortex,  extracted using diffusion spectrum imaging and tractography   \cite{Hagmann2008},  \cite{honey2009predicting} (see  \cite{2791},  \cite{arsiwalla2015network}, \cite{arsiwalla2015connectomics} for neurodynamical models used on this network). Compared to a randomly re-wired network, it was seen that the particular topology of the human brain generates greater information complexity for all allowed couplings associated to the network's attractor states, as well as to its non-stationary dynamical states \cite{xda2016global}.  However, the formulation of  $\Phi$ is not specific to biological systems and can equally well be applied to artificial dynamical systems and serves as a measure of their computational or information processing complexity (which we interpret as cognitive complexity or awareness in biological agents).

\section{Synthetic Consciousness: Insights from Synthetic Biology and Artificial Intelligence }

Although our understanding of natural systems can be strongly constrained by 
experimental limitations, the potential for exploring synthetic counterparts provides a unique 
research window. It has been suggested that artificial simulations, {\em in silico}   implementations  and engineered alternatives 
can actually be much needed to understand the origins of evolutionary dynamics, including cognitive transitions \cite{sole2016synthetic}.  What can be achieved in relation to 
consciousness from artificial agents? 

Within the context of non-cognitive phenomena,  synthetic biology provides a valuable example of the  classes of relevant questions that 
can be answered. Examples are the possibility of creating living  systems from non-living chemistry, generating multicellular assemblies, 
creating synthetic organoids  or even artificial immune systems. Here advanced genetic engineering techniques along with a systems 
view of biology had been able to move beyond standard design principles provided by evolution.  Examples of this are  new genetic codes
 with extra genetic letters in the alphabet that have been designed and successfully inherited \cite{artdna2014}, synthetic protocells with replicative 
potential \cite{sole2007synthetic} and even whole synthetic chromosomes that have defined  a novel bacterium species  \cite{mingen2016}.  
Ongoing research has also revealed the potential for creating cognitive networks of interacting microorganisms capable of 
displaying collective intelligence \cite{sole2016sci}.


Of course, the criteria for consciousness, as stated in sections above, are not even remotely satisfied by any of these synthetic  systems. They either have some limited form of intelligence or life but not yet both. Nevertheless, there have been some noteworthy recent developments in these areas.  AlphaGo's feat in beating the top human Go champion was remarkable for a couple of reasons. Unlike Chess, possible combinations in Go run into the  millions and when played using a timer any brute-force algorithm trying to scan the entire search space would simply run out of computational capacity or time. Hence, an efficient  pattern recognition algorithm  was crucial to the development of AlphaGo, where using deep reinforcement learning the system was trained on a large number of games after which it  was made to play itself over and over again (this aspect of playing itself is akin to  training via social interactions as described  later on)  while reinforcing successful sequence of plays through the weights of its deep neural networks \cite{silver2016}. Most interestingly, it played counterintuitive moves that shocked the best human players and the sole game of the series that Lee Sedol, the human champion won out of five, itself was only possible after he himself adopted a brilliant counterintuitive strategy. Thus, AlphaGo demonstrates a form of domain-specific intelligence. In contrast, biological awareness spans across domains.  Moreover, AlphaGo is not equipped with any form of arousal mechanisms coupled to its computational capabilities.  

The same can be said for other state-of-the-art AI systems including deep convolutional neural networks, or deep recurrent networks. Both these latter architectures were inspired from Hubel and Weisel's groundbreaking work on the coding properties of the visual system, which led to the realization of a hierarchical processing architecture    \cite{hubel1962receptive}.  Today deep convolutional networks are widely used for image classification \cite{ciresan2011flexible} and recurrent neural networks for speech recognition \cite{sak2014long}, among countless other applications.  The current interest of deep learning has been anticipated in computational neuroscience using objective functions from which physiologically plausible perceptual hierarchies can be learnt  \cite{wyss2006model}. Recent developments have advanced this by virtue of larger data sets and more computational power. For example,  there have been attempts to build biologically-plausible models of learning in the visual cortex using recurrent neural networks    \cite{liao2016bridging}. In  summary,  deep architectures have made remarkable progress in domain-specific AI. 
  
However, asking whether a machine can be conscious in exactly the same way  that a human is, is similar to asking whether a submarine can swim. It just does it differently. If the goal of a system is to learn and solve complex tasks close to human performance or better, current machines are already doing that in specific domains.  However, these machines are still far from learning and solving problems in generic domains and in  ways that would couple its problem-solving capabilities to its autonomous survival drives. On the other hand, neither have any of the synthetic life systems discussed above been used to build architectures   with complex computing or cognitive capabilities. Nevertheless, this does suggest that a  future synthesis between artificial life forms and AI could be evaluated using  homologous scales of  consciousness to the ones currently used  for biological life forms. This form of synthetic consciousness, if based on a life form with different survival drives/mechanisms and non-human forms of intelligence or computation, would also likely lead to non-human behavioral outcomes. 

These phenomenological considerations suggest at least two generic types of complexities to label states of consciousness, those associated to computational/informational capabilities and those referring to autonomous functioning. In the following section, we argue for a third complexity type, necessary to build the morphospace of consciousness, namely, social complexity. 

\section{The Function of Consciousness:  Insights from Cognitive Robotics }
 Based on insights from cognitive robotics, this section takes a functional perspective on consciousness    \cite{conscious12016},  \cite{arsiwalla2017consciousness},  \cite{herreros2016forward},  \cite{moulin2017embodied},  \cite{verschure2016synthetic},    \cite{verschure2014and}  eventually interpreting it as a game-theoretic strategy.  In \cite{verschure2016synthetic}, it was suggested that rather than being the  problem itself, consciousness might in fact be a solution to the problem of autonomous goal-oriented  action  with intentionality, when agents are faced with a multi-agent social environment. The latter was formulated as the \emph{H5W problem}.

\subsection{The H5W Problem }

What does an agent operating in a social world need to do in order to optimize its fitness? It needs to perceive the world, to act and, through time, to understand the consequences of its actions so it can start to reason about its goals and how to achieve them. This requires building a representation of the world grounded on the agent's own sensorimotor history and use that to reason and act. It will witness a scene of agents, including itself, and objects interacting in various manners, times and places. This comprises the six fundamental problems that the agent is faced with, together referred to as the H5W  problem  \cite{verschure2016synthetic}:  In order to act in the physical world an  agent needs to determine a behavioral procedure to achieve a goal state; that is, it has to answer the HOW of action. In turn this requires the agent to: (a)  
Define the motivation for action in terms of its needs, drives and goals, that is, the WHY of action; (b) Determine knowledge of  objects it needs to act upon and their affordances in the world, pertaining to the above goals, that is, the WHAT of action; (c) Determine the location of these objects, the spatial configuration of the task domain and the location of the self, that is, the WHERE of action; (d) Determine the sequencing and timing of action relative to dynamics of the world and self, that is, the WHEN of action; and (e) Estimate hidden mental states of other agents when action requires cooperation or competition, that is, the WHO of action.

While the first four of the above questions suffices for generating simple goal-oriented behaviors, the last of the Ws (the WHO) is of particular significance as it involves intentionality, in the sense of estimating the future course of action of other agents based on their social behaviors and psychological states.  However, because mental states of other agents that are predictive of their actions are hidden, they can at best be inferred from incomplete sensory data such as location, posture, vocalizations, social salience, etc.  As a result the acting agent faces the challenge to univocally assess, in a deluge of sensory data those exteroceptive and interoceptive states that are relevant to ongoing and future action and therefore has to deal with the ensuing credit assignment problem in order to optimize its own actions. Furthermore, this results in a reciprocity of behavioral dynamics, where the agent is now acting on a social and dynamical world that is in turn acting upon itself. It was proposed in \cite{verschure2016synthetic}  that consciousness is associated to  the ability of an agent to maintain a transient and autonomous memory of the virtualized agent-environment interaction, that captures the hidden states of the external world, in particular, the intentional states of other agents and the norms that they implicitly convey through their actions. 

\subsection{Social Game Theory  }

Hence, the function of consciousness is to enable an acting agent to solve its H5W problem while being engaged in social cooperation and competition   with other agents, who are trying to solve their own H5W problem in a world with limited resources. This leads our discussion precisely within the setting of social game theory.  In a scenario with only a small number of other agents, a given agent might use statistical learning approaches to learn and classify behaviors of the few others agents in that game. For example, multiple robots interacting to learn naming conventions of perceptual aspects of the world \cite{steels2003}. Here the multi-agent interaction has to be embodied so that one agent can interpret which specific perceptual aspect the other agent is referring to (by pointing at objects)  \cite{steels2012}.  

Another example is the emergence of signaling languages  in sender-receiver games based on replicator dynamics described by David Lewis in 1969 in his seminal work, Convention \cite{david1969convention}, \cite{lewis2008convention}. However,   in both these examples, strategies that evolutionarily succeed when only few players are involved, are no longer optimal in the event of an explosion in the number of players  \cite{hofbauer2008feasibility}. Likewise in a social environment comprising a large number of agents trying to solve the H5W problem, machine learning strategies for reward and punishment valuations may soon become computationally unfeasible for an agent's processing capacities and memory storage. Therefore, for a large population to sustain itself in an evolutionary game involving complex forms of cooperation and competition would require strategies other than simple machine learning algorithms. One such strategy involves modeling and inferring intentional states of itself and that of other agents. Emotion-driven flight or fight responses depend on such intentional inferences and so do higher-order psychological drives. The mechanisms of consciousness enable such strategies, whereas, contemporary  AI systems such as AlphaGo do not possess such capabilities.
  
In summary, interpreting consciousness as a game-theoretic strategy highlights the role of complex social behaviors inevitable for survival in a multi-agent world. From an evolutionary standpoint, social behaviors result from generations of cooperation-competition games, with natural selection filtering out unfavourable strategies. Presumably, winning strategies were  eventually encoded as anatomical mechanisms, such as emotional responses. The complexity of these behaviors depends on the ability of an agent to make complex social inferences. This suggests a third dimension in the morphospace of consciousness (shown in figure \ref{fig2}), namely, social complexity, which serves as a measure of an agent's social intelligence.


\begin{table}[htp]
	\begin{centering}
		\begin{tabular}{|c|c|c|c|}
\hline
  &   ${\cal C}_{Autonomous}$   & ${\cal C}_{Computational}$   & ${\cal C}_{Social}$    \\
\hline  
Substrate & Organism, nervous  & Cognitive systems &  Interacting population \\
  &system, bots &(brains, microprocessors)& of agents \\
 \hline  
Parts  & Sensors, actuators, &Neurons, transistors& Individual agents  \\  
& signalling cascades &&  \\   
\hline 
Emergence  & Self-regulated   &Problem solving & Signaling systems,  \\
& real-time behavior & capabilities & language, social norms,   \\
&&& conventions, art, \\  
&&&   science, culture \\
\hline
\end{tabular}
		\caption{The three complexity types along with their respective substrates, components and emergent properties.   }
		\label{table1}
	\end{centering}
\end{table}

\section{A Morphospace of Consciousness }

As evident from our discussions above, consciousness research draws  insights from a variety of disciplines such as  clinical neuroscience, synthetic biology, artificial intelligence, evolutionary biology and cognitive robotics. Taken together, this suggests at least three complexity types (see figure 2 and table 1)  that can be associated to consciousness: autonomous complexity, computational complexity and social complexity.  As a generic definition of a system's complexity ${\cal C}$, we define   
\begin{eqnarray}
{\cal C} = {\cal I}_{substrate} - \sum_{\{parts\}}   {\cal I}_{part}
\end{eqnarray}
which is a measure of information generated by the dynamics of a system as a whole (${\cal I}_{substrate}$) minus the sum of that generated by its parts. While this is similar in spirit to integrated information discussed in an  earlier section,  it is generically defined for specifying substrate-specific complexity. This provides a general framework that includes the possibility of  several different types of complexity, among which,  ${\cal C}_{Autonomous}$,  ${\cal C}_{Computational}$  and    ${\cal C}_{Social}$ will be relevant for labelling states of consciousness.  

Autonomous complexity ${\cal C}_{Autonomous}$ measures the complexity of autonomous actions. In eukaryotes, autonomous action refers to arousal mechanisms resulting from coordinated nervous system activity; in prokaryotes, autonomous action refers to reactive behaviours such as chemotaxis, stress responses to temperature, toxins, mechanical damage, etc., all of these resulting from coordinated cellular signalling processes; in robotics, autonomous action refers to homeostatic mechanisms driving reactive behaviors. Therefore,  autonomous complexity is the information generated by the collective dynamics of the complex system driving autonomous actions, over the information generated by a (hypothetical) uncoordinated copy of this system.  On the other hand, computational complexity ${\cal C}_{Computational}$  refers to the ability of an agent to integrate  information over space and time across computational or cognitive tasks.  

In complex biological systems, this complexity is typically associated to neural processes, in artificial computational systems, it  refers to microprocessor signaling. The distinction between ${\cal C}_{Computational}$ and  ${\cal C}_{Autonomous}$ is specified by the tasks that they refer to, rather than the specific substrate.  
Finally, social complexity ${\cal C}_{Social}$  refers to the information generated by the population as a whole, during the course of social interactions,  over the information generated additively by individual agents of the population.  Unlike  ${\cal C}_{Autonomous}$ or ${\cal C}_{Computational}$,  ${\cal C}_{Social}$  is not assigned to an individual, but rather to a specific   population (its own species) with which the individual has been interacting. Nonetheless, as discussed above, these interactions are believed to have contributed to the consciousness of an individual on an evolutionary time-scale, by way of social games.  Note that  ${\cal C}_{Social}$ as defined here, does not refer  to group consciousness (we shall  discuss that in the following section).

%
%

\begin{figure}[h]
\centering
\includegraphics[width=1.0\linewidth]{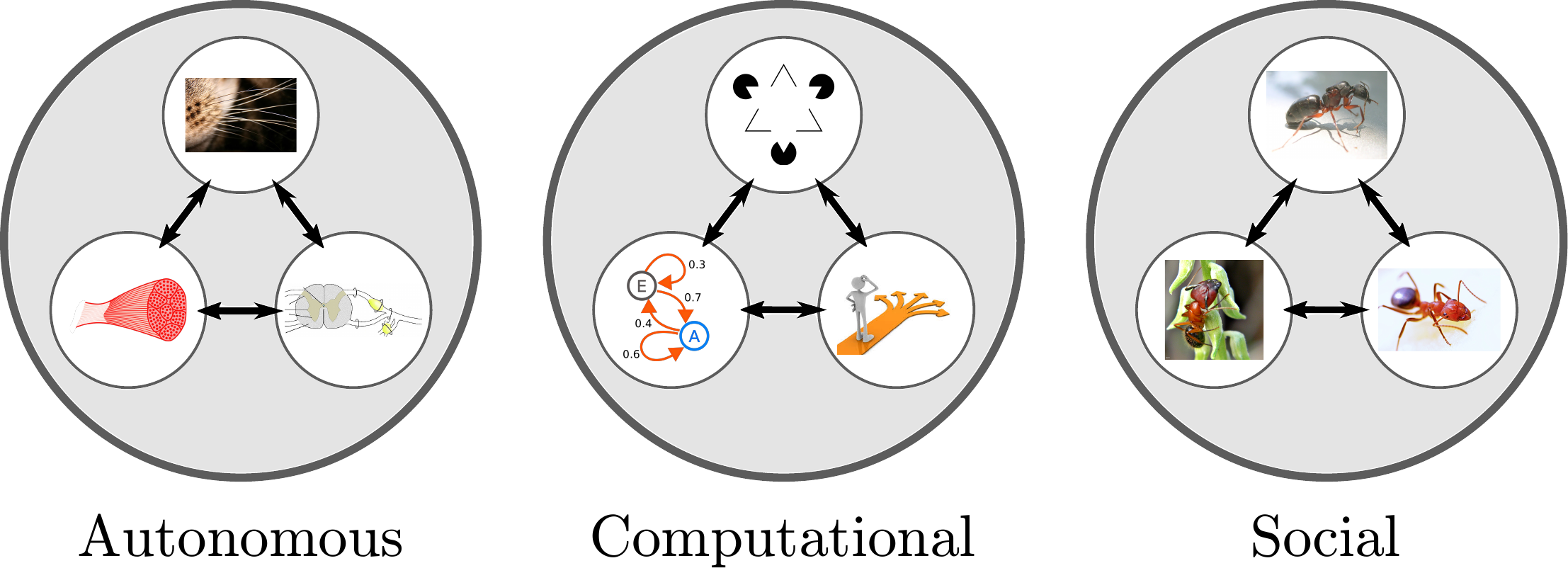}
\caption{{\bf  Schematic representation of autonomous, computational and social complexity.}  Each complexity measure is illustrated as a whole (the large circles) constituted of its parts (the inner circles), their interactions (the arrows) and the emerging properties resulting from these interactions (the inner space within the large circles, in light grey). Autonomous complexity (left) refers to the collective phenomena resulting from the interactions between typical components of reactive behavior such as sensors (illustrated by whiskers in the top inner circle), actuators (illustrated by a muscle in the bottom-left inner circle) and low-level sensorimotor coupling (illustrated by a spinal cord in the bottom-right inner circle).  Computational complexity is associated to higher-level cognitive processes such as visual perception (top inner circle), planning (bottom-left inner circle) or decision making (bottom-right inner circle). Social complexity is associated to interactions between individuals of a population, such as a queen ant (top inner circle), a worker ant (bottom-left inner circle) and a soldier ant (bottom-right inner circle). }
\label{fig3}
\end{figure}

\begin{figure}[h]
\centering
\includegraphics[width=0.8\linewidth]{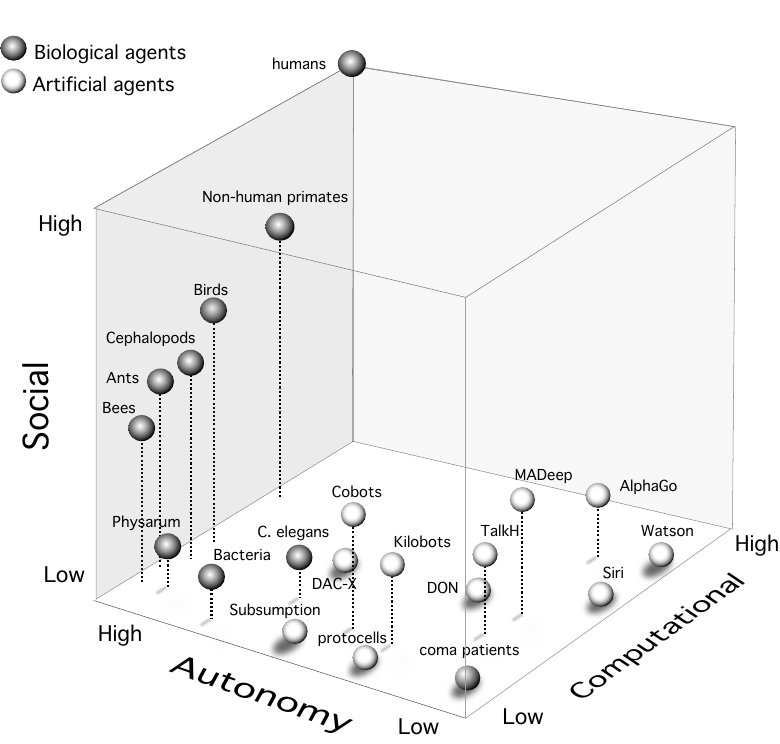}
\caption{{\bf  Morphospace of consciousness.}  Autonomous, computational and social complexity constitute the three axes of the consciousness morphospace.  Human consciousness is used as a reference in one corner of the space. Current AI implementations cluster together in the high computation, low autonomy and low social complexity regime, while multi-agent cognitive robotics cluster around low computational, but moderate autonomous and social complexities.  Abbreviated legends: MADeep (multi-agent deep reinforcement system) \cite{Tampuu2015};  TalkH (talking heads)  \cite{steels2012};  DQN  (deep Q-learning)  \cite{mnih2015human};  DAC-X (distributed adaptive control)  \cite{maffei2015embodied}, CoBot (cockroach robot)  \cite{halloy2007social},  Kilobot (swarm robot)  \cite{rubenstein2014}, 
Subsumption (mobile robot architecture)  \cite{Brooks1986}. }
\label{fig2}
\end{figure}

Using these definitions for the three types of complexities, we construct the following morphospace in figure \ref{fig2}.    
While this space is only a first attempt at constructing the space of prospective conscious systems, the precise coordinates of various systems within this morphospace might change due to the rapid pace of new and developing technologies, but we expect the relative locations of each example to remain the same. We start with the human brain, which is taken as the benchmark in this space, defining a limit case 
located at one upper corner with highest scores on all the three axes. She/he can perform computational tasks across a variety of domains such as making logical inference, planning an optimal path in a complex environment or dealing with recursive problems and hence leads with respect to computational complexity due to these cross-domain capabilities. On the social axis, human social interactions have resulted in the emergence of language, music, art, culture or socio-political systems. Other biological entities such as  non-human primates  \cite{borjon2016arousal}, \cite{de2016apes} or social insects would score lower on the social and computational axis than humans.  Additionally, other species of vertebrates such as some types of birds and cephalopods have been  shown to exhibit complex behavior and possess sophisticated nervous systems. These two groups have actually being enormously useful 
in the search for animal consciousness   \cite{edelman2009animal},    \cite{emery2004mentality}.


Current AI systems such as IBM Watson \cite{high2012era},  AlphaGo \cite{silver2016}, DQNs \cite{mnih2015human}  and Siri \cite{aron2011innovative}  are powerful computing systems over a narrow set of domains, but in their current form  they do not show general  intelligence, that is, the capacity to independently interact with the world and successfully perform different tasks in different domains \cite{Legg2007}, or as proposed by Allen Newell, the capability with which anything can become a task \cite{Newell1973}. These AI systems  are still clustered high on the computational axis, but lower than humans (due to domain-specificity).  Also they score low on autonomy and social complexity. Synthetic forms of life such as protocells show some levels of arousal, reacting to chemicals and stressors, but currently show minimal capabilities for computation or adaptation and no interactions with other agents  \cite{protocell2015}.  

Interest in the field of collective robotics has led to the rise of machines where emergent macro-properties, e.g. coordination (KiloBot \cite{rubenstein2014}, Multi-Agent Deep Network  \cite{Tampuu2015}) or shared semantic conventions (Talking Heads \cite{steels2012}) self-organize out of  multi-agent interactions. These systems are designed to display simple forms of navigation, object-detection, etc., while interacting with other agents performing the same task. However, they show lower social and autonomous complexity than most biological agents, and being embodied, they currently score lower on computational complexity than heavy-powered AI systems such as IBM Watson and AlphaGo.  Notice also that a large region in the central zone of the morphospace in figure \ref{fig2}  is  suspiciously  vacant.   A similar observation was made in \cite{olle2016morphospace} in the context of the morphospace of synthetic organs and organoids. In both cases,  such an observation points towards  new classes of  future machines.  In the following section, we discuss two possible manifestations  of  future conscious systems.    

An important use of the morphospace within evolutionary biology is related to the actual occupation of this space by 
the different solutions. In the previous figure it is possible to appreciate that a large part of the space is empty. 
Along with the biological case studies, the set of artificial solutions remain (so far) in a lower part of the cube, 
thus indicating the small relevance played by the social context. Social interactions have instead played a leading role 
in shaping the minds of the organisms close to the left wall involving high autonomy and sociality. Here the 
nature of the social axis changes among case studies (as well as the underlying computational complexity). But 
the filter of evolution has a major impact: social insects are enormously resilient to many environmental challenges 
because no individual is more relevant than another. Redundancy and distributed computation define then as 
cognitive assemblies. In contrast, cognitively complex organisms equipped with brains and exhibiting cooperative 
behavior have been evolved to live together with others.

\section{Other Embodiments of Consciousness }

The three dimensional morphospace discussed above provides us with a framework to also identify other types of complex systems whose levels of computational, autonomous and social complexity might be sufficient to answer the H5W of consciousness?  This suggests at least two other embodiments of  future conscious systems (based on the same functional criteria as above).

\subsection{Group Consciousness   }

In a sense biological consciousness itself can be thought of as a collective phenomenon where individual cells making up an organism are themselves not considered to be conscious (with respect to the three complexity measures defining the morphospace), even though the organism as a whole is. But what happens when the system itself is not localized?  We postulate  group  consciousness as an  extension of the above idea to composite or distributed systems that display levels of computational, autonomous and social complexity that are  sufficient to answer the H5W problem.  Note that, as per this specification of group consciousness, the group itself is treated as one entity. Hence, social complexity now refers to the interactions of this group with other  similar  groups.   

This bears some resemblance to the notion of collective intelligence, which is a widely studied phenomenon in complex systems ranging from ant colonies \cite{dorigo2007swarm}, to a swarm of robots (the Kilobot in   \cite{rubenstein2014} and the CoRobot in \cite{halloy2007social}), to social networks \cite{goleman2007social}. But these are generally not regarded as conscious systems. As a whole they are not considered to be life forms with survival drives that compete or cooperate with other similar agents. However, these considerations begin to get blurred at least during transient epochs when collective survival comes under threat. For example, when a bee colony comes under attack by hornets, collectively it demonstrates a prototypical survival drive, similar to lower-order organisms.  Other examples of such behaviors have also been studied in the context of group interactions in humans, where social sensitivity, cooperation and diversity have been shown to correlate with the collective intelligence of the group \cite{woolley2010evidence}. Following this, the notion of collective intentionality has been discussed in  \cite{huebner2013macrocognition}.  More recently,   \cite{detm2017}  have applied   integrated information $\Phi$ to group interactions, suggesting a new kind of group consciousness.  While it is known that $\Phi$ in adapting agents increases with fitness \cite{edlund2011integrated}, one can ask a similar question for an entire group: what processes (evolutionary games, learning, etc.) enable an increase in all three complexity types  for an entire group such that it can solve the H5W problem while cooperating or competing with other groups?  


\subsection{Simulated Consciousness }

Our discussions on complexity also suggest another type of consciousness, namely, simulated consciousness, wherein embodied virtual agents in a simulated reality   interact with other virtual agents, while satisfying the complexity bounds that enable them to answer the H5W questions within the simulation. In this case,  consciousness is strictly confined to the simulated environment. The agents cannot perceive or communicate with entities outside of the simulation but satisfy all the criteria we have discussed above within the simulation.    How these embodied virtual agents could acquire consciousness is not yet known. Presumably by evolving across multiple generations of agents that adapt and learn to optimize fitness conditions. It is also not clear what precise traits or mechanisms would have to be coded into the simulation (as initializations or priors) in order to enable consciousness to evolve. The point here is simply that the same criteria that we have identified with consciousness in biological or synthetic agents in the physical world, could in principle be admitted by agents within a simulation and confined to their interactions within that simulation. This has parallels to the notion of "Machine Consciousness" discussed in \cite{reggia2013rise}, which proposes that neural processes leading to consciousness might be realizable as a machine simulation (it even goes further to claim that computer systems might someday be able to emulate consciousness).  At the moment, these are all open challenges in AI and consciousness research. Examples of studies discussing embodied virtual agents can be found in the work of \cite{cassell2000embodied}  and  \cite{burden2009deploying}. More recent implementations of embodied virtual agents have been using  gaming technology, such as the Minecraft platform  \cite{aluru2015minecraft},     \cite{johnson2016malmo}.

\section{Consciousness and General Intelligence}

What do our discussions concerning consciousness have to say about theories of general intelligence? The idea that consciousness resides in select regions of a morphospace, that is constructed from function-specific types of complexity, has implications for any theory of artificial general intelligence. Namely, it suggests a specific decomposition of general intelligence into complementary types. In psychology, distinct manifestations of human intelligence have been discussed in the context of Howard Gardner's theory of multiple intelligences   \cite{gardner2011frames}.  Here we want to understand how the dimensions of our morphospace help group different types of intelligences. This works as follows\footnote{We thank Carlos E. Perez for bringing this point to our attention.  A  discussion about how Gardner's intelligence types may be realized in machines  using deep learning can be found in his recent book \cite{perez2017deep}.}.  The autonomous axis reflects adaptive intelligence found in biological organisms. This encapsulates Gardner's kinesthetic, musical and spatial intelligence (some of these also require computational complexity).  The computational axis refers to  recognition, planning and decision-making capabilities that we find in computers as well as in humans. These are tasks involving logical deduction or inference. Hence, this complexity refers to those types of intelligences that require computational capabilities, such as logical reasoning, linguistic intelligence, etc.  The third axis of the morphospace,   social complexity, relates to social capabilities required for interacting with other agents. This refers to interpersonal and introspective intelligence, in Gardner's terms. These types of intelligences are also associated to the evolution of language, social conventions and culture. Then there are also other types of intelligences described in Gardner's theory such as naturalistic and pedagogical intelligence, which involve a composition of social and computational complexity. 

As described above, the defining dimensions of our morphospace account for all of the multiple types of intelligences proposed by Gardner.  Taking these intelligence (or complexity) types into account, while building artificially intelligent machines, elucidates the wide spectrum of problems that future AI could potentially address.  In the light of both, Gardner's theory and Newell's criteria, our morphospace in fact suggests, that consciousness as we know it, manifests as a specific form of integrated multiple intelligence. Note that one ought to be careful {\em not} to claim that consciousness '{\em  is}' general intelligence. Following William James, in cognitive psychology, consciousness is rather seen as a process \cite{james1892stream}. We claim that this process constitutes mechanisms and phenomenology that realizes an integration of specific types of intelligences and their associated complexities in such a way so as to meet survival goals. Intelligence, on the other hand, can be thought of as a task-specific capability, that by itself is not necessarily tied to any existential pressures \cite{xda2018phy}.  However, currently we have yet to understand how many of the intelligence types mentioned above, especially the non-computational ones \cite{arsiwalla2018brains},  can even be realized individually,  let alone understanding the mechanisms that lead to their integration. Nonetheless, given the myriad of recent  advances in human-machine interactions, a complexity-based conceptualization of consciousness provides a practical and quantitative framework for studying ways in which interactions with machines might enhance our joint complexities and competences.


\section{Societal and  Ethical Considerations }

No discussion on conscious machines is complete without the very important issue of ethics.  Both, the societal impact and ethical considerations of any form of advanced   machine, especially conscious machines, for obvious reasons, constitutes a very serious issue. For example, the impact of medical nanobots for removing tumors, attacking viruses or non-surgical organ reconstruction has the potential to change medicine forever. Or AI systems to clear pollutants from the atmosphere or the rivers are absolutely essential for some of the biggest problems that humanity faces. However, as discussed above, purely increasing the performance of a machine along the computational axis will not constitute consciousness as along as these capabilities are not accessible by the system to autonomously regulate or enhance its survival drives. On the other hand, whenever the latter is indeed made possible, issues of societal interactions of machines with humans and the ecosystem, becomes an imminent ethical responsibility. It becomes important to understand the kind of cooperation-competition dynamics that a futuristic human society will face. Early stages of designing such machines are probably the best times to regulate their future impact on society.  This analogy might not be surprising to any parent that has a child. Hence, a serious effort  towards  understanding the evolution of complex social traits is crucial alongside engineering advances required for the development of  these    systems.

\section{Discussion }

The objective of this article was to bring together diverse ideas from neuroscience, AI, synthetic biology and robotics, that have recently been converging towards the science of consciousness.  Following progress in these fields, we have attempted to generalize the applicability of current clinical scales of consciousness to synthetic systems. In particular,  starting from clinical measures of consciousness that calibrate awareness and wakefulness in patients, we have investigated how contemporary AI systems and synthetically engineered organisms compare on homologous measures.  Awareness and wakefulness  can be abstracted to computational and autonomous complexity respectively.  Additionally, using insights from cognitive robotics, we have discussed functions that consciousness serves in nature, and argued that consciousness manifests as an evolutionary game-theoretic strategy. This made the case for a third type of complexity necessary to describe consciousness, namely, social complexity. These three complexity types allow us to represent both, biological and synthetic systems in a common morphospace. 

A morphospace is a useful construct to study systems-level properties of complex systems based on information-theoretic measures. The three complexity types comprising the morphospace described here, are representative of biological as well as synthetic complex systems. Using this morphospace, we have shown how various biological organisms including  bacteria, bees, C. elegans, primates and humans compare to artificially intelligent  systems such as deep networks, multi-agent systems, social robots,  intelligent assistants such as Siri and computational systems as IBM's  Watson.  Besides biological and synthetic consciousness, these considerations also suggest other possible manifestations of consciousness, namely, group consciousness and simulated consciousness, each based on distinct embodiment.  

In our discussion, social complexity was crucial for constructing  the  morphospace. Social interactions play an important role in regulating many cognitive and adaptive behaviors in both, natural and artificial systems  \cite{verschure2014and}.  In  \cite{verschure2016synthetic},  it has been suggested  that complex social interactions may have evolutionarily served as a trigger for consciousness. What is however not known  is whether there are specific lower bounds on the scales of each of the stated    complexity types, that an agent must cross in order to attain a given level of consciousness.  Certainly, from developmental biology  we know that both humans (and many higher-order animals) undergo extensive  periods of cognitive and social learning from infancy to maturation. These phases  of social and cognitive training are necessary   for development of cognitive abilities leading to levels of consciousness attained in adulthood.    


Even though we may be far from understanding all the engineering principles required to build conscious machines, a complexity-based comparison between biological and artificial systems reveals interesting insights. For example, current AI systems using deep learning tend to cluster along the computational complexity axis of the morphospace, whereas synthetically engineered life forms group closer along the autonomous complexity axis. On the other hand, biologically conscious agents are distributed in regions of the morphospace corresponding to relatively high complexity along all the three axes (which suggests necessary, if not sufficient, conditions for consciousness).  In terms of Newell's criteria,  excluding those that refer exclusively to human-specific traits (language, symbolic reasoning), the remainder are completely satisfied by all agents located in the high complexity region (of all three axes) of the consciousness morphospace. In contrast, current AI or synthetic systems do not check-out on this list. Though in 1994 Newell was not explicitly referring to consciousness, it is remarkable to note how those ideas to formulate theories of cognition and intelligence seem to reconcile with current ideas of consciousness. One could summarize the crux of Newell's criteria as referring to   agents displaying autonomous behaviors with cross-domain problem-solving capabilities, which can be decomposed to (at least) the three complexity classes discussed in this paper.   

This perspective on consciousness opens several possibilities for future work.  For instance, it may be interesting to further refine the morphospace described here.  In particular, computational complexity itself may involve several sub-types involving learning, adaptation, acquiring sensorimotor representations, etc, all of which are relevant for cognitive robotics  \cite{verschure2003environmentally}.  Another question arising out of our discussion is whether the emergence of consciousness in a multi-agent social environment  can be identified as a Nash equilibrium of a cooperation-competition game.  In a game where say two species attain consciousness, the population pay-offs in cooperation and competition   between them are likely to reach one of possible equilibria due to the recursive nature of intentional inferences, where an agent attempts to infer the inferences of other agents about its own intentions.  Multi-agent models might offer a viable  approach to test ideas such as these.

\section*{Acknowledgments} 
We thank Riccardo Zucca and Sytse Wierenga for help with graphics. This work has been supported by the European Research Council's CDAC project: "The Role of Consciousness in Adaptive Behavior: A Combined Empirical, Computational and Robot based Approach" (ERC-2013- ADG 341196).  RS acknowledges the support of the Secretaria d'Universitats i Recerca del Departament d'Economia i Coneixement de la Generalitat de Catalunya, the Botin Foundation, by Banco Santander through its Santander Universities Global Division and by the Santa Fe Institute.

\section*{References}
\bibliographystyle{splncs03}
\bibliography{test2}

\end{document}